\documentstyle[12pt]{article}
\input epsf

\newcommand{\beq}{\begin{equation}}
\newcommand{\eeq}{\end{equation}}
\newcommand{\beqn}{\begin{eqnarray}}
\newcommand{\eeqn}{\end{eqnarray}}
\newcommand{\gtrsim}{\stackrel{>}{\sim}}
\newcommand{\lessim}{\stackrel{<}{\sim}}

\newcommand{\dst}{\displaystyle}
\newcommand{\bm}{\boldmath}
\newcommand{\fr}[2]{\frac{{\dst #1}}{{\dst #2}}}
\newcommand{\eq}[1]{eq. (\ref{#1})}

\newcommand{\ptr}{\mbox{$p_{\bot}\,$}}
\newcommand{\ptrs}{\mbox{$p_{\bot}^2\,$}}
\newcommand{\ggam}{\mbox{$\gamma\gamma\,$}}

\title{Semihard diffractive production of neutral mesons by
off shell photons and the range of pQCD validity\thanks{
The earlier version of this paper was distributed as preprint
[1]}}
\author{I.F.~Ginzburg$^{a)}$ , D.Yu.~Ivanov$^{a)}$ and
V.G.~Serbo$^{b)}$}
\date{{\it $^{a)}$Institute of Mathematics, 630090 Novosibirsk,
Russia\\ $^{b)}$Novosibirsk State University, 630090 Novosibirsk,
Russia}}

\begin{document}

\maketitle

\begin{abstract}
We study the dependence on photon virtuality $Q^2$ for the semihard
quasi--elastic photoproduction of neutral vector mesons on a quark,
gluon or real photon (at $s\gg p_{\bot}^2,\;Q^2; \; p_{\bot}^2
\gg \mu^2 \approx (0.3$ GeV)$^2$). To this end we calculate the
corresponding amplitudes (in an analytical form) in the lowest
nontrivial approximation of the perturbative QCD. The amplitudes
for the production of mesons consisting of light quarks vary
very rapidly with the photon virtuality near $Q^2=0$.

We estimate the bound of the pQCD validity region for such
processes. For the process with mass shell photon the obtained
bound is very high, and this region seems beyond opportunities
of real experiment. This bound decreases fast with the increase
of $Q^2$, and we expect that the virtual photoproduction at HERA
provides the opportunity to test the pQCD results. The signature
of this region is discussed.
\end{abstract}

\section{Introduction}

The diffractive photoproduction of neutral vector mesons $V$
is studied in many theoretical [1--12] and experimental
[13--15] papers.

In this paper we study this photoproduction on quark, gluon or
other photon, initiated by off shell photon $\gamma^*$ (with
virtuality $Q^2$):
\beq
\gamma^* q\to Vq,\quad \gamma ^* g\to Vg;\qquad
\gamma^*\gamma \to  V V^{\prime}.\label{0}
\eeq
in the region of parameters where the perturbative QCD (pQCD)
validity is beyond doubts:
\beq
s\gg p_{\bot}^2,\;Q^2;\;\;\; p_{\bot}^2\gg\mu^2\qquad
(\mu\approx 0.2\div 0.3 \mbox { GeV}).
\label{range}
\eeq
The transverse momentum of produced meson relative to collision
axis \ptr is small as compared to energy but it is large as
compared to QCD scale $\mu \approx 0.3$ GeV.

These reactions can be studied in the photoproduction on proton
with the rapidity gap $\eta>\eta_0$ between produced meson and
other produced hadrons X. {\em We denote such processes as
diffractive photoproduction.} The cross section of
the process $\gamma^* p \to V X$ with rapidity gap is related with
that for the
photoproduction on quark and gluon via the well known
relation\footnote{ $G(x,t)$ and $q(x,t)$ are the gluon and quark
densities in proton.}:
\beqn
\frac{d\sigma (\gamma^* p\to VX)}{dtdx}&=&
\sum_f\left(q(x,t)+\bar q(x,t)\right)
\frac{d\sigma(\gamma^* q\to Vq)}{dt}+\nonumber\\
&&G(x,t)\frac{d\sigma(\gamma^* G\to VG)}{dt};\quad
x>\fr{4\ptrs}{s}\cosh^2\fr{\eta_0}{2}.\nonumber
\eeqn

The photon--photon collisions of the discussed type can be
also studied at the future photon colliders \cite{GKST}.

To estimate the bounds of the pQCD validity region, we simulate
the nonperturbative effects near these bounds by the specific
model. Its idea is to use the pQCD equations, in which quark mass
is considered as a parameter (which is near the constituent quark
mass). We also use this model for the qualitive description of
some phenomena below this bound.

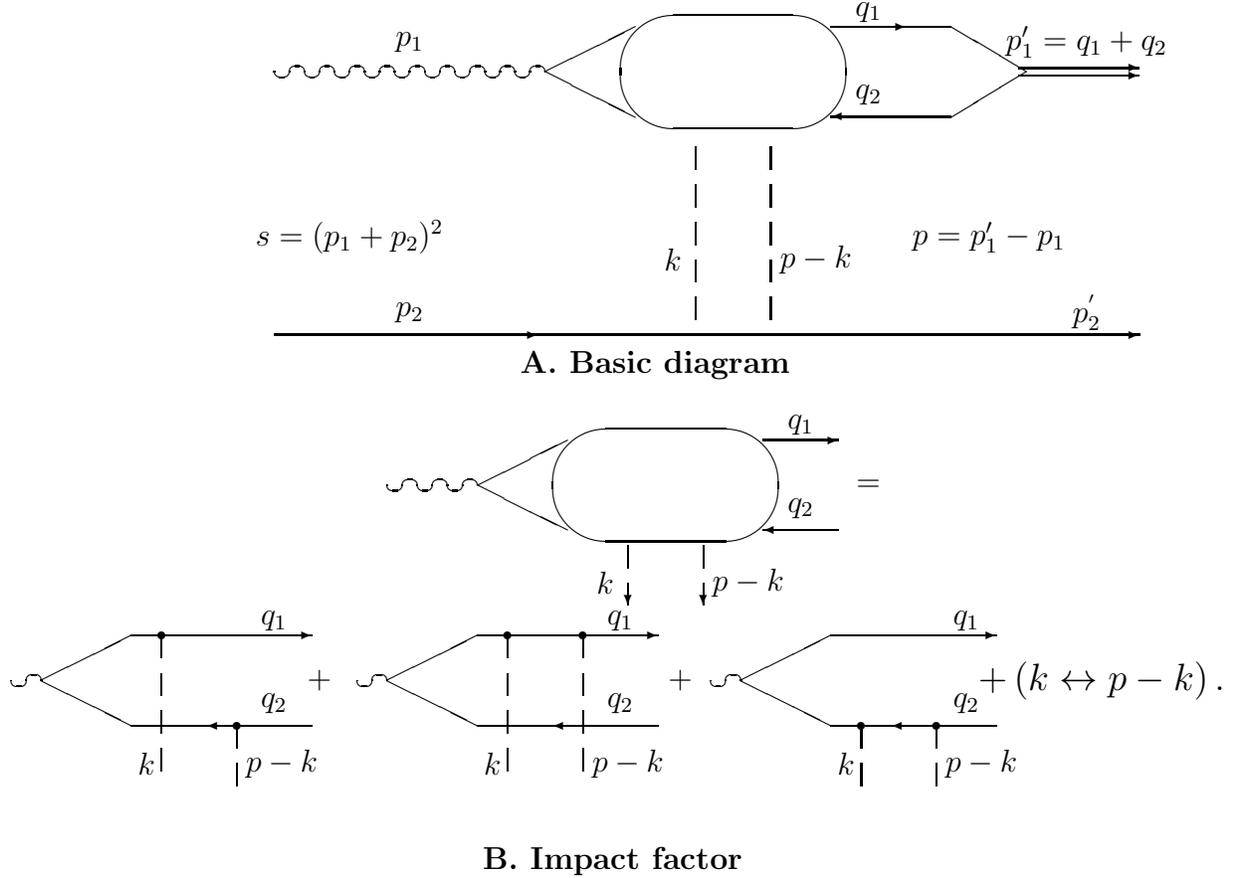
\begin{figure}[hbt]
\begin{center}
\unitlength 1mm
\begin{picture}(160,115)(0,-10)
\multiput(36,100)(4,0){9}{\oval(2,1.5)[b]}
\multiput(38,100)(4,0){9}{\oval(2,1.5)[t]}
\put(71,100){\line(2,1){12}}
\put(71,100){\line(2,-1){12}}
\put(96,100){\oval(30,15)}
\put(109,106){\vector(1,0){10}}
\put(125,94){\vector(-1,0){16}}
\put(119,106){\line(1,0){6}}
\put(114,108){\makebox(0,0){$ q_1$}}
\put(114,97){\makebox(0,0){$ q_2$}}
\put(53,104){\makebox(0,0){$ p_1$}}
\put(143,104){\makebox(0,0){$ p'_1=q_1+q_2$}}
\put(135,100){\line(-5,3){10}}
\put(135,100){\line(-5,-3){10}}
\put(134,100.5){\vector(1,0){16}}
\put(134,99.5){\vector(1,0){16}}
\put(35,65){\vector(1,0){115}}
\put(60,65){\vector(1,0){10}}
\multiput(91,67)(0,5){5}{\line(0,1){3}}
\multiput(101,67)(0,5){5}{\line(0,1){3}}
\put(53,68){\makebox(0,0){$ p_2$}}
\put(143,68){\makebox(0,0){$ p_{2}^{'}$}}

\put(88,75){\makebox(0,0){$ k$}}
\put(107,75){\makebox(0,0){$ p-k$}}
\put(130,78){\makebox(0,0){$ p=p'_1-p_1$}}
\put(45,78){\makebox(0,0){$ s=(p_1+p_2)^2$}}

\put(85,61){\makebox(0,0){ \bf A. Basic diagram}}
\multiput(51,45)(4,0){3}{\oval(2,1.5)[b]}
\multiput(53,45)(4,0){3}{\oval(2,1.5)[t]}
\put(62,45){\line(2,1){12}}
\put(62,45){\line(2,-1){12}}
\put(87,45){\oval(30,15)}
\put(100,51){\vector(1,0){10}}
\put(110,39){\vector(-1,0){10}}
\put(105,53){\makebox(0,0){$ q_1$}}
\put(105,42){\makebox(0,0){$ q_2$}}
\put(79,32){\makebox(0,0){$ k$}}
\put(98,32){\makebox(0,0){$ p-k$}}
\put(82,37){\line(0,-1){3}}
\put(82,32){\vector(0,-1){3}}
\put(92,37){\line(0,-1){3}}
\put(92,32){\vector(0,-1){3}}
\put(114,45){\makebox(0,0){$ =$}}

\put(1,19){\oval(2,1.5)[b]}
\put(3,19){\oval(2,1.5)[t]}
\put(4,19){\line(2,1){12}}
\put(4,19){\line(2,-1){12}}
\put(16,25){\vector(1,0){24}}
\put(40,13){\vector(-1,0){14}}
\put(16,13){\line(1,0){10}}
\put(35,27){\makebox(0,0){$ q_1$}}
\put(35,16){\makebox(0,0){$ q_2$}}
\multiput(20,25)(0,-5){4}{\line(0,-1){3}}
\multiput(30,13)(0,-5){2}{\line(0,-1){3}}
\put(18,8){\makebox(0,0){$ k$}}
\put(36,8){\makebox(0,0){$ p-k$}}
\put(20,25){\circle*{1}}
\put(30,13){\circle*{1}}
\put(41,19){\makebox(0,0){$ +$}}

\put(47,19){\oval(2,1.5)[b]}
\put(49,19){\oval(2,1.5)[t]}
\put(50,19){\line(2,1){12}}
\put(50,19){\line(2,-1){12}}
\put(62,25){\vector(1,0){24}}
\put(86,13){\vector(-1,0){14}}
\put(62,13){\line(1,0){10}}
\put(81,27){\makebox(0,0){$ q_1$}}
\put(81,16){\makebox(0,0){$ q_2$}}
\multiput(76,25)(0,-5){4}{\line(0,-1){3}}
\multiput(66,25)(0,-5){4}{\line(0,-1){3}}
\put(64,8){\makebox(0,0){$ k$}}
\put(82,8){\makebox(0,0){$ p-k$}}
\put(76,25){\circle*{1}}
\put(66,25){\circle*{1}}
\put(89,19){\makebox(0,0){$ +$}}

\put(94,19){\oval(2,1.5)[b]}
\put(96,19){\oval(2,1.5)[t]}
\put(97,19){\line(2,1){12}}
\put(97,19){\line(2,-1){12}}
\put(109,25){\vector(1,0){22}}
\put(131,13){\vector(-1,0){14}}
\put(109,13){\line(1,0){8}}
\put(127,27){\makebox(0,0){$ q_1$}}
\put(127,16){\makebox(0,0){$ q_2$}}
\multiput(123,13)(0,-5){2}{\line(0,-1){3}}
\multiput(113,13)(0,-5){2}{\line(0,-1){3}}
\put(111,8){\makebox(0,0){$ k$}}
\put(129,8){\makebox(0,0){$ p-k$}}
\put(123,13){\circle*{1}}
\put(113,13){\circle*{1}}

\put(145,19){\makebox(0,0){\large$+\left( k \leftrightarrow
p-k\right). $}}

\put(80,-5){\makebox(0,0){\bf B. Impact factor}}

\end{picture}

\caption{\em Photoproduction of vector meson on a quark
(two--gluon exchange).}

\end{center}
\label{fig1}
\end{figure}

Our efforts are focused on the problems: {\em What are main
features of $Q^2$-dependence in these processes within pQCD,
without any phenomenological hypotheses?  What are the bounds for
pQCD validity at the description of diffractive processes?} To
this end we restrict our consideration by calculation in the
lowest nontrivial approximation of pQCD --- two--gluon exchange
for production of vector mesons (Fig. \ref{fig1}). The  obtained results
provide opportunity to discuss the relation between the
point--like and hadron--like components of a photon in the
discussed reactions as well.

The known for us papers, treated similar problems, are discussed
briefly in the last section.

The study of pQCD validity in the discussed processes is on line
with that in other exclusive processes. The relation between
perturbative and nonperturbative contributions in their
description and the bounds for pQCD validity are discussed widely
(see refs. \cite{I,Br} and references therein).  The advantage of
processes considered is the possibility to study this subject by
two probes simultaneously --- via investigation of dependence on
both the produced meson transverse momentum $p_{\bot}$ and the
photon virtuality $Q^2$.

\section{Basic relations}

The process discussed can be described as two stage one. At the
first stage,  photon fragments into a $q\bar q$ pair, the quarks
with energies $\varepsilon_i$ move along the photon momentum.
Their transverse momenta are relatively small and the total
energy of quark's pair is close to the energy of the photon,
$\varepsilon_1+\varepsilon_2\approx E$. This first stage
describes also the processes with production of jet-like system
(both resolved for two quark jets and unresolved one) with
rapidity gap.

At the second stage then quarks are glued into meson.

The basic kinematical notations are presented in Fig. \ref{fig1}. We
denote also the virtuality of photon by $Q^2\equiv -p_{1}^2 >0$,
the quark mass --- by $m$, the transverse momentum of produced
meson (relative to the collision axis) --- by ${\bf p_{\bot}}$.
For the description of the photon fragmentation into quarks, we
use quark spinors $u_1=u(q_1)$ and $u_2=u(-q_2)$. The relative
motion of the quark and antiquark is described by a variable $\xi
$ that is the ratio:

\beq
\xi =\fr{2\left(q_1-q_2\right)p_2}{s}\equiv \fr
{\varepsilon _1-\varepsilon _2}{E} ;\qquad -1\leq \xi \leq 1.
\nonumber
\eeq
Next, we denote
\beq
{\bf n} = {{\bf p}_{\bot} \over \mid {\bf p}_{\bot}\mid};\quad
\delta = {2m\over p_{\bot}};\quad u={Q^2\over {\bf
p}_{\bot}^2};\; v= \delta^2 + (1-\xi^2)u= {4m^2+(1-\xi^2)Q^2
\over {\bf p}_{\bot}^2}.
\label{u}
\eeq
Besides, $e= (0,{\bf e},0)$ and $e_V=(0,{\bf e}_V,0)$ are the
polarization vectors of the transverse photon and the
transversely polarized vector meson (in the state with helicity
$\lambda =\pm 1$).

As usually, $ \alpha _s= g^2/4\pi,\;\alpha =e^2/4\pi =1/137,\;Q_q
e$ is the quark charge, $N=3$ is the number of colors.\\

{\large\bf Impact representation}\\

The amplitude of the process in the lowest nontrivial order of
pQCD is described by diagrams of Fig. \ref{fig1} (with accuracy
$\sim\ptrs/s,\; Q^2/s $). Just as in refs. \cite{GPS1,GPS2,GIv},
the sum of these diagrams is transformed with the same accuracy
to an integral over the gluon transverse momentum --- {\em the
impact representation}:
\beq
M_{\gamma ^*q\to Vq}=is\int \fr {J_{\gamma ^*V}({\bf
k}_{\bot},{\bf p}_{\bot})\; J_{qq}(-{\bf k}_{\bot},-{\bf
p}_{\bot})} {{\bf k}^2_{\bot}\; ({\bf k}_{\bot}- {\bf
p}_{\bot})^2}\fr{d^2k_{\bot}}{(2\pi)^2}.
\label{4}
\eeq

Impact factors $J_{\gamma ^* V}$ and $J_{qq}$ correspond to the
upper and lower blocks in Fig. \ref{fig1}. They are $s$-independent. The
entire dependence on the photon virtuality is concentrated in
the impact factor $J_{\gamma^* V}$. For colorless exchange
impact factors include factors $\delta_{ab}$, where $a$ and $b$
are the color indices of the exchanged gluons.

The impact factors for the transition between two colorless
states vanish when the gluon momenta tend to zero \cite{GPS1}.
This general property takes place independent on validity of
perturbation theory. (In the coordinate space this property can
be treated as zero's color charge of object. This property is
named "dipole shielding", "quark coherence", etc.). In our case
this property is written as
\beq
J_{\gamma^* V}({\bf k}_{\bot},{\bf p}_{\bot})\to 0\;\mbox{ at }\;
\left\{
\begin{array}{cr}
{\bf k}_{\bot}&\to 0,\\ {\bf (p-k)}_{\bot}& \to 0.
\end{array}\right.
\label{dipol}
\eeq

The derivation of impact representation and impact factors
repeats in main features that given in Appendices to ref.
\cite{GPS1} (see refs. \cite{LFr,ChWu} also) with two
variations.

First, Sudakov variables are introduced precisely for the
reaction with "massive" collided particles. All momenta are
decomposed over "almost light-like" vectors combined from initial
ones: $ p'_1 = p_1-(p^2_1/s) p_2=p_1 +(Q^2/s) p_2\;$, $\;p'_2=p_2
-(p^2_2/ s) p_1 $, and in the plane perpendicular to them.
Simple calculations with these vectors show that the masses
squared in the denominators of quark propagators become more
"heavy":
\beq
m^2\to m^2+ Q^2(1-\xi^2)/4.\label{mq}
\eeq

Second, the impact factors for these photons are different for
the production by transverse ($\gamma^*_T$) and scalar
(or longitudinal) ($\gamma^*_S$) off shell photons.

When consider the entire pQCD series in the leading log
approximation (LLA), one can use the method of calculation from
ref. \cite{MT}. In this case the impact representation
transforms to the form (see Fig. \ref{fig2}):
\beq
M_{\gamma ^*q\to Vq}=is\int J_{\gamma ^*V}({\bf k}_{\bot},{\bf
p}_{\bot})\; J_{qq}(-{\bf k'}_{\bot},-{\bf p}_{\bot}) {\cal
P}(s;{\bf\ptr, k_{\bot},k'_{\bot}})
\fr{d^2 k_{\bot}d^2 k'_{\bot}}{(2\pi)^4}. \label{pomerimp}
\eeq

\begin{figure}[hbt]
\begin{center}
\unitlength 1mm
\begin{picture}(120,50)
\multiput(1,45)(4,0){9}{\oval(2,1.5)[b]}
\multiput(3,45)(4,0){9}{\oval(2,1.5)[t]}
\put(36,45){\line(2,1){12}}
\put(36,45){\line(2,-1){12}}
\put(61,45){\oval(30,15)}
\put(74,51){\vector(1,0){10}}
\put(84,51){\vector(1,0){6}}
\put(90,39){\vector(-1,0){16}}
\put(0,0){\vector(1,0){115}}
\put(25,0){\vector(1,0){10}}
\put(18,49){\makebox(0,0){$ p_1$}}
\put(18,3){\makebox(0,0){$ p_2$}}

\put(53,32){\makebox(0,0){$ k$}}
\put(72,32){\makebox(0,0){$ p-k$}}
\put(56,37){\line(0,-1){3}}
\put(56,32){\vector(0,-1){3}}
\multiput(56,7)(0,5){2}{\line(0,1){3}}
\put(56,5){\vector(0,-1){3}}
\put(53,8){\makebox(0,0){$k'$}}
\put(66,37){\line(0,-1){3}}
\put(66,32){\vector(0,-1){3}}
\multiput(66,7)(0,5){2}{\line(0,1){3}}
\put(66,5){\vector(0,-1){3}}
\put(72,8){\makebox(0,0){$ p-k'$}}
\put(61,22){\oval(16,16)}
\put(61,22){\makebox(0,0){\large$\cal P$}}
\put(100,45){\line(-5,3){10}}
\put(100,45){\line(-5,-3){10}}
\put(99,45.5){\vector(1,0){16}}
\put(99,44.5){\vector(1,0){16}}
\put(108,42){\makebox(0,0){\large V}}

\end{picture}

\caption{}

\end{center}
\label{fig2}

\end{figure}

The discussed lowest nontrivial approximation of pQCD (\ref{4})
corresponds to
$$
{\cal P}({\bf \ptr,k_{\bot},k'_{\bot}})=
\fr{(2\pi)^2\delta({\bf k}_{\bot}-{\bf k}'_{\bot})}
{{\bf k}^2_{\bot}\; ({\bf k}_{\bot}- {\bf p}_{\bot})^2}
\delta_{aa'}\delta_{bb'}.
$$

The kernel ${\cal P}$ of this equation relates to the
perturbative Pomeron (pP). The impact factors, obtained in the
basic approximation of pQCD, are also valid for the description
of process in LLA for both mass shell and off shell photons
until virtuality $Q^2$ is not too large.

In its asymptotic form the amplitude (\ref{pomerimp}) is
Regge--like:
\beq
M=is G_{\gamma^* V}(\ptr,Q^2)\cdot K(s/\ptrs)\cdot G_{q q}.\nonumber
\eeq
The kernel $K$ is pP itself, it is obtained in refs. \cite{Lip}.
Each vertex $G_{\gamma^* V}$ is the convolution of our $\gamma^*
V$ impact factor with some standard factor from ${\cal P}$. The
corresponding integration is similar to that in our case, and
the $Q^2$ dependence near mass shell is rougjly the same. The
detailed calculation of $Q^2$ dependence here is absent now. The
calculations of ref. \cite{Iv} shows that this Regge--like form
is valid at large enough $\eta\approx\ln(s/\ptrs) \gtrsim 3$ for
real photons. At smaller values of rapidity gap $\eta $ the
lowest order calculations related to approximation of Fig. \ref{fig1}
seems more adequate for the description of data.

The theoretical and experimental study of pP is of great
interest, since this object should be common for different
reactions and it is sensitive to the inner structure of pQCD. In
particular, it is important to test BFKL \cite{Lip} predictions
about pP in its pure form without mixing with large distance
(soft) effects.\\

{\large\bf Quark and gluon impact factors}\\

The impact factor for the colorless transitions $q\to q$ (from
\cite{GPS1}) and $g\to g$ are similar:
\beq
J_{qq}=g^2\;{\delta_{ab} \over 2N};\quad J_{gg}=-g^2\delta _{ab}
\;{N\over N^2-1}.
\label{14}
\eeq

The helicity and color state of the quark or gluon target are
conserved in these vertices.

{\large\em Gluon dominance.} The relations (\ref{14}) shows that
the cross section for the photoproduction of vector meson on a
gluon is about 5 times larger than that on a quark:
\beq
d\sigma _{\gamma ^* g \to Vg} = \left(\fr{2N^2}{N^2-1}\right)^2
d\sigma _{\gamma ^*q \to Vq}={81\over 16 } d\sigma _{\gamma ^*
q \to Vq}.\label{GlQ}
\eeq
It means that the photoproduction of vector meson on proton with
a rapidity gap can be used for study of the gluon content of a
proton.

Having in mind these facts, we will present the formulae for the
photoproduction on the quark for the definiteness.

\newpage

{\large\bf Impact factor $J_{\gamma^* q\bar q}$}\\

{\large\em The impact factor for a transverse photon} has the
same form as for the on shell photon \cite{GPS1} but with the
replacement (\ref{mq}) in denominators:
\beq
J_{\gamma ^*_T q\bar q}=eQ_qg^2 \;{\delta_{ab}\over 2N}\;{\bar u}_1
\left[mR(m){\hat e}-\left(1+\xi\right){\bf P}(m){\bf e}-\hat P(m)\hat e
\right]  \; {{\hat p}_2 \over s}\;u_2. \label{5}
\eeq

Here transverse vector $P(m)=(0,{\bf P}(m),0)$ and scalar $R(m)$ are:
\beqn
{\bf P}(m)&=&
\left[\fr{{\bf q}_{1\bot}}{ {\bf q}^2_{1\bot}+m^2+(1-\xi
^2)Q^2/4} \;+ \;\fr {{\bf k}_{\bot}-{\bf q}_{1\bot}}{ ({\bf
k}_{\bot}-{\bf q}_{1\bot})^2 +m^2+(1-\xi ^2)Q^2/4}\right] - \label{7}\\
&&-\left[{\bf q}_{1\bot}\leftrightarrow {\bf q}_{2\bot}\right];\nonumber\\
&&\nonumber\\
R(m)&=&\left[{1 \over {\bf q}^2_{1\bot}+m^2+(1-\xi ^2)Q^2/4} \;- \;
{1 \over ({\bf k}_{\bot}-{\bf q}_{1\bot})^2 +m^2+(1-\xi
^2)Q^2/4}\right] +\label{7a}\\
&& +\left[{\bf q}_{1\bot}\leftrightarrow {\bf q}_{2\bot}\right]. \nonumber
\eeqn

To describe {\large \em the impact factor for a scalar photon},
it is necessary to know the polarization vector of scalar photon
$e_S$. Taking into account the gauge invariance,one can use a
reduced form of this vector $ e_S=2\sqrt{Q^2}\; (p'_2/ s)$ in
our kinematical region. Then the calculations similar to those
for $T$ photon result in:
\beq
J_{\gamma ^*_S q\bar q}=-eQ_qg^2{
\delta_{ab}\over 2N}\;  \fr{1-\xi ^2}{2}\sqrt{Q^2}\;R(m)\;
{\bar u}_1 \; {{\hat p}_2 \over s}\; u_2.
\label{6}
\eeq

It is easily seen that these impact factors obey eq. (\ref{dipol}).\\

{\large\bf Impact factors for a meson production}\\

To produce a meson, the relative transverse momenta of quarks
should be small ($\stackrel {<}{\sim} \mu$). With our accuracy
($\mu^2/\ptrs\ll 1$) the transverse momenta of quarks relative to
the collision axis are proportional to their energies
$\varepsilon_i$, i.e.
$$
{\bf q}_{1\bot}=\fr{1}{2}(1+\xi ){\bf p}_{\bot},\quad {\bf
q}_{2\bot}=\fr{1}{2}(1-\xi ){\bf p}_{\bot},\quad
\varepsilon_{1,2}=\fr{1}{2}(1\pm \xi )E.
$$

The $q \bar q \to V$ transition is described, as usual (see
\cite{LeBr}), by change of product $\bar u_1...u_2$ for the
meson wave function $\varphi_V (\xi)$:
\beq
Q_q\bar u_1\dots u_2\to \fr{Q_V}{4N}\int \limits_{-1}^{1}d\xi
\left\{\begin{array}{ccc}
f_V^L \varphi_V^L(\xi )\mbox {Tr}\left( \dots\hat p_3\right)&
\mbox{ for }V_L\\
f_V^T \varphi_V^T(\xi ) \mbox {Tr}\left(\dots
\hat e^*_V\hat p_3\right) & \mbox{ for } V_T.
\end{array}\right.\label{2}
\eeq
(The trace over vector and color indices is assumed). The
quantity $Q_V$ relates to the quark charges in the meson $V$.
The specific forms for these wave functions is given in \eq{3a},
\eq{3}. We use the coupling
constants from refs. \cite{ChAZh,BaGr}:

\begin{center}
\begin{tabular}{|c|c|c|c|c|c|c|c|c|}\hline
  &$\rho^0$&$\omega$&$\phi$&$\Psi$&
$\Psi'$&$\Upsilon$&$\Upsilon'$&$\Upsilon''$\\ \hline
&&&&&&&&\\$f_V$, GeV
&0.21&0.21&0.23&0.38&0.28&0.66&0.49&0.42\\ &&&&&&&&\\ $Q_V$
&$1/\sqrt {2}$&$1/(3 \sqrt{2})$ &1/3&2/3&2/3&1/3&1/3&1/3\\
\hline
\end{tabular}
\end{center}
\vspace{0.3cm}

The impact factors $J_{\gamma^* V}$ are obtained by substitution
of eq. (\ref{2}) into eqs. (\ref{5}),(\ref{6}):

a) For {\em a transverse photon} we have two opportunities:
\beq
J_{\gamma _T ^* V}({\bf k}_{\bot},{\bf p}_{\bot})= {1\over 2}eQ_V
g^2 {\delta _{ab} \over 2N}\; \int
\limits^{1}_{-1}d\xi \left\{
\begin{array}{ll}
(-f^L_V)\varphi ^L_V(\xi )\;\xi \;({\bf P} {\bf e}) & \mbox{ for
} V_L\\ f^T_V\varphi ^T_V(\xi )\;mR\;({\bf e}{\bf e}^*_V) & \mbox
{ for } V_T.\end{array}\right.\nonumber
\eeq

b) {\em A scalar photon} produces a longitudinal vector meson
only:
\beq
J_{\gamma _S ^* V _L}({\bf k}_{\bot},{\bf p}_{\bot})= -{1\over
2}eQ_V g^2 {\delta _{ab} \over 2N}\; \int
\limits^{1}_{-1}d\xi f^L_V \varphi ^L_V (\xi)
{1-\xi ^2 \over 2}\; \sqrt{Q^2}\;R.\nonumber
\eeq
Below we neglect difference between $\varphi^L$ and $\varphi^T$,
$f^L$ and $f^T$.

It is useful to introduce dimensionless vector ${\bf r}$ via
equation $ {\bf k}_{\bot}= ({\bf r}+{\bf n}) p_{\bot}/2$. Then
the above impact factors acquire the forms:
\beqn
J_{\gamma ^* V}& = & eQ _V g^2 {\delta _{ab} \over 2N} \;{ f_V
\over \mid p_{\bot}\mid}
\left\{\begin{array}{ll}
\left({\bf e}{\bf F}_{T\to V_L}\right)& \mbox{ for T-photon}\to
\mbox { meson }V_L\\
\delta\left({\bf e}{\bf e} ^*_V\right) F_{T\to V_T}&
\mbox{ for T-photon}\to\mbox { meson }V_T\\
F_{S\to V_L} & \mbox{ for S-photon}\to\mbox { meson }V_L.
\end{array}\right.\label{12}\\
{\bf F}_{T\to V_L}& =& - \int \limits^{1}_{-1}d\xi \varphi _V(\xi
)\cdot \xi
\left\{\left[\fr{(1+\xi){\bf n}}{v+(1+\xi)^2 }+
\fr{{\bf r}-{\bf n}\xi}{v+\left({\bf r}-{\bf n}\xi\right)^2}
\right]-\left[\xi \leftrightarrow - \xi\right]\right\};\label{12a}\\
F_{T\to V_{T}}&=&\int \limits^{1}_{-1}\varphi_V(\xi) d\xi\cdot {\cal R};\quad
F_{S\to V_{L}}=-\int \limits^{1}_{-1}d\xi \sqrt{u}(1-\xi^2)\varphi_V(\xi)
\cdot{\cal R}\label{12b}\\
{\cal R}&=& \left[\fr{1}{v+(1+\xi)^2 }-\fr{1}{v+\left({\bf r}+
{\bf n}\xi\right)^2}\right]+\left[\xi \leftrightarrow -
\xi\right].\nonumber
\eeqn

\section {The neutral vector meson photoproduction
on a quark or gluon}

To calculate amplitudes under interest we substitute these impact
factors into eq. (\ref{4}). The result for the meson production
on a quark is
\beq
M_{\gamma ^*q\to Vq}=i {eQ_V g^4 \over \pi}\; {s f_V\over
|p_{\bot}| ^3} \;{N^2-1\over N^2}\; \left\{
\begin{array}{ll}
\left({\bf e}{\bf n}\right)I_{T\to V_L}  &\mbox{ for T-photon}
\to\mbox { meson }V_L \\
\left({\bf e}{\bf e}_V^{\ast} \right)
\delta \cdot I_{T\to V_T}  & \mbox{ for T-photon}\to
\mbox { meson }V_T \\
I_{S\to V_L} &\mbox{ for S-photon}\to\mbox { meson }V_L
\end{array}\right.\label{15}
\eeq
with
\beq
I_a\,=\,{1\over 4\pi}
\int \fr{\displaystyle {F_a({\bf r},{\bf n})}}
{\displaystyle {\left({\bf r}-{\bf n}\right)^2\left({\bf r}+ {\bf
n}\right)^2}} d^2r\equiv
\int \limits^{1}_{-1}d\xi\;\varphi_{V}(\xi)\Phi_a(\xi)
\;\;\left(a= T\to V_L,\;T\to V_T,\; S\to V_L\right).\label{T0}
\eeq
(For $a= T\to V_L$ the quantity $({\bf n F}_{T\to V_L})$ is
used.)

Just as in refs. \cite{GPS1,GPS2}, we integrated over component
of vector ${\bf r}$ along ${\bf n}$ using residues. Last
integration is trivial (but bulky).  Then the quantities in eq.
(\ref{15}) get the form
\beqn
\Phi_{T\to V_L}&=&{\xi \over 4(1-\xi^2-v )}
\left [{(1+\xi )^2-v \over (1+\xi )^2+v}
\ln{{(1+\xi )^2+v \over 2\sqrt{v}}}  -
{(1-\xi )^2-v \over (1-\xi )^2+v}
\ln{{(1-\xi )^2+v \over 2\sqrt{v}}}
\right ] \nonumber\\
\Phi_{T\to V_T}&=&{1 \over 2(1-\xi^2-v )}
\left [{(1+\xi ) \over (1+\xi )^2+v}
\ln{{(1+\xi )^2+v \over 2\sqrt{v}}}  +
{(1-\xi ) \over (1-\xi )^2+v}
\ln{{(1-\xi )^2+v \over 2\sqrt{v}}}
\right ] \nonumber\\
\Phi_{S\to V_L}&=&-\sqrt{u}(1-\xi^2)\Phi_{T\to V_T}
\ .
\label{16}
\eeqn

We will discuss below the scale of $Q^2$--dependence for cross
sections. Let us define this scale $\Lambda^2$ by equation
\beq
I_T(\Lambda^2) ={1\over 2}I_T(Q^2=0).
\eeq

\subsection{Production of mesons consisting of heavy quarks}

The calculations and results below differ for the production of
mesons consisting of heavy or light quarks. We begin with a more
simple case of mesons consisting of heavy quarks ($J/\Psi$ or
$\Upsilon$). It seems more clean since the large quark mass
suppresses nonperturbative effects. The results are similar in
main features to those obtained in ref.\cite{GPS2} for the mass
shell photons. We will speak below about the $J/\Psi$ meson
photoproduction for definiteness.

For the wave function of discussed mesons we use the usual main
approximation:
\beq
\varphi(\xi)\;=\;\delta(\xi).\label{3a}
\eeq
With this wave function the impact factor $T\to V_L$ (for
production of longitudinally polarized vector meson by transverse
photon) vanishes. The deviation from the simple form (\ref{3a})
can be described by the quantity $<\xi^2>=\int\xi^2\varphi(\xi)\,
d\xi \sim 0.1$.

Let us begin with the {\large\bf photoproduction on a quark.} The
main results for the transverse photon coincide with those for
real photons \cite{GPS2} with the replacement $\delta^2 \to \nu$.
Finally, in eq. (\ref{15})
\beqn
I_{T\to \Psi_T}&=& {1\over 2(\nu^2-1)} L(\nu);\quad L(\nu)=
\ln{{(1+\nu )^2 \over 4\nu } };\;I_{S\to \Psi_L}=-\sqrt{u}
I_{T\to \Psi_T};\label{PsiT}\\
I_{T\to \Psi _L}&=& {<\xi ^2>\over (1+\nu)^2}\left[1-{\nu
\over \nu-1} L(\nu)\right];\qquad \nu =
(4m^2+Q^2)/{\bf p}_{\bot}^2 .\label{Psi0}
\eeqn

Therefore, the helicity conserves in these reactions. {\em The
transverse photon produces mainly transverse $\Psi$}. The
admixture of longitudinally polarized $\Psi$ is $\sim(<\xi^2>)^2
\sim 0.01 $ \cite{GPS2}.  {\em Scalar photon produce
longitudinal $\Psi _L$ only.} The ratio of amplitude with
production of transverse $J/\Psi$ by $T$-photons to that with
production of longitudinal $J/\Psi$ by $S$-photons is
$\sqrt{4m_c^2/ Q^2}\,$, it is independent on $p_{\bot}$. At
$Q^2\;> 4m^2$ the dominant polarization becomes longitudinal.

The largest amplitudes $\gamma^*_T q\to \Psi_T q$ and $\gamma^*_S
q\to \Psi_L q$ vanish at $p^2_\bot = 4m_c^2+Q^2$ (or $\nu =1$).
These zeroes shift strongly due to $<\xi^2>$ corrections. At the
higher values of $p_{\bot}$ these cross sections are small (cf.
ref. \cite{GPS2}).

The shape of both main amplitudes is determined by the single
function $I_{T\to \Psi_T}$. One can see that the scale $\Lambda^2
$ of $Q^2$ dependence increases from the natural value $4m_c^2/2$
at small $p_{\bot}$ to $\sim p_{\bot}^2/10$ at large enough
$p_{\bot}$.

The similar calculations give us the amplitudes for the {\large
\bf\bm production of two mesons $V',\,V$ consisting of heavy
quarks in \ggam collision}. We consider the production of both
identical and different mesons by real or virtual photons. In
particular, the collision of the virtual photon with the real
one is described by two nonzero amplitudes, the first ---
for the production by $T$-photon and the second --- for the
production by $S$-photon (these amplitudes are finite at $p_\bot
\to 0$):
\beqn
M_{\gamma ^*_T \gamma \to V'_T V_T}&=& {is \over p_{\bot} ^4}
\;{e^2 g^4 \over \pi} Q_V Q_{V'} f_V f_{V'} {N^2-1\over N^2}\;
({\bf e}_1 {\bf e}_{V'}^*)\;({\bf e}_2 {\bf e}_V^*)\;\delta
\delta '\cdot I_{V'V} ; \nonumber\\
M_{\gamma ^*_S \gamma \to V'_L V_T }&=& -{is \over p_{\bot} ^4}
\;{e^2 g^4 \over \pi} Q_V Q_{V'} f_V f_{V'} {N^2-1\over N^2}
\;({\bf e}_2 {\bf e} _{V}^*)\;\sqrt{u'} \delta\cdot I_{V'V} ;
\label{V'V}\\
&&I_{V'V}={1\over u'} \left[ {L(\delta^2)\over (\nu'+1)(\delta
-1)}\; - \; {L(\nu')\over (\nu'-1)(\delta +1)}\right].
\nonumber
\eeqn
Here $\nu'$ corresponds to meson produced by off shell photon,
and $\delta$ --- to on shell one.

\subsection{Production of mesons consisting of light quarks on
a quark or gluon}

The impact factor $J_{\gamma ^*_T V_T}$ contains factor $\delta
= 2m/p_{\bot}$. Therefore, {\bf in the range of pQCD validity
(at large enough \ptr) the transverse photons produce mesons
consisting of light quarks in the states with helicity $0$ only}
for any polarization of an initial photon\footnote{ It is in
contrast with well known fact, that at small $p_{\bot}$ helicity
conserves mainly, i.e. vector mesons are transversely
polarized.}. It is due to the chiral nature of perturbative
couplings in the massless limit.

We write in this section for shortness $I_{T}$ instead
of $I_{T\to V_L}$ and
\beq
I_S\equiv I_{S\to V_L}(u)\,\equiv\,- {2\sqrt{u}\over
1+u}(I_{T}\,+\,U).\label{IST}
\eeq

We use the wave functions of mesons consisting of light quarks in
the form \cite{ChAZh}:
\beq
\varphi _V(\xi )=\fr{3}{4} \left( 1-\xi
^2\right)\left(1-\fr{1}{5}b_V+b_V\xi^2\right).\label{3}
\eeq
Coefficient $b_V$ tends to 0 slowly with growth of $\ptrs \quad
(b_{\rho}=b_{\omega}=1.5;\;\; b_{\phi}=0$ at $\ptr\approx 1$ GeV).

For the asymptotical wave function ($b_V = 0$) we obtain:
\beqn
I_T(u)&\equiv& I_0 (u)=\fr{\displaystyle 3}{\displaystyle
8(1-u)^3} \left[ 2+10u-u\left(\fr{\displaystyle
1+u}{\displaystyle 1-u}\right) \left(\ln^2{\fr{\displaystyle
1}{\displaystyle u}}+6\ln{ \fr{\displaystyle 1}{\displaystyle
u}}\right)\right]; \nonumber\\ U(u)&\equiv&U_0
(u)\,=\,\fr{\displaystyle 3}{\displaystyle 8(1-u)}
\left( 2- \fr{\displaystyle 1+u}{\displaystyle 1-u}
\ln{\fr{\displaystyle 1}{\displaystyle u}}
\right).\label{19}
\eeqn
For the case $b_V\neq 0$ the more complicated expressions are
obtained:
\beqn
I_{T}(u)&=&I_0 (u)\left [ 1- {b_V\over 5}+b_V \left( {1+u\over
1-u}\right)^2 \right] + \nonumber\\
&+&{ b_V\over
12(1-u)^4}\left[-(3-10u)(1-u)+16u(u^2+5u+1)U_0
(u)\right];\label{21}\\
U(u)&=&U_0 (u)+\,\fr{b_V}{60(1-u)^2}\left[
5(1-u) \,+\,8(1+8u+u^2) U_0 (u)\right].
\nonumber
\eeqn
These expressions are regular at $u=1$. At $u=0$ (real
photoproduction) $I_T\;=\;{3\over 4}(1+{7\over 15} b_V)$
\cite{GPS1}.

\begin{figure}
\epsfxsize=15cm
\centerline{\epsffile{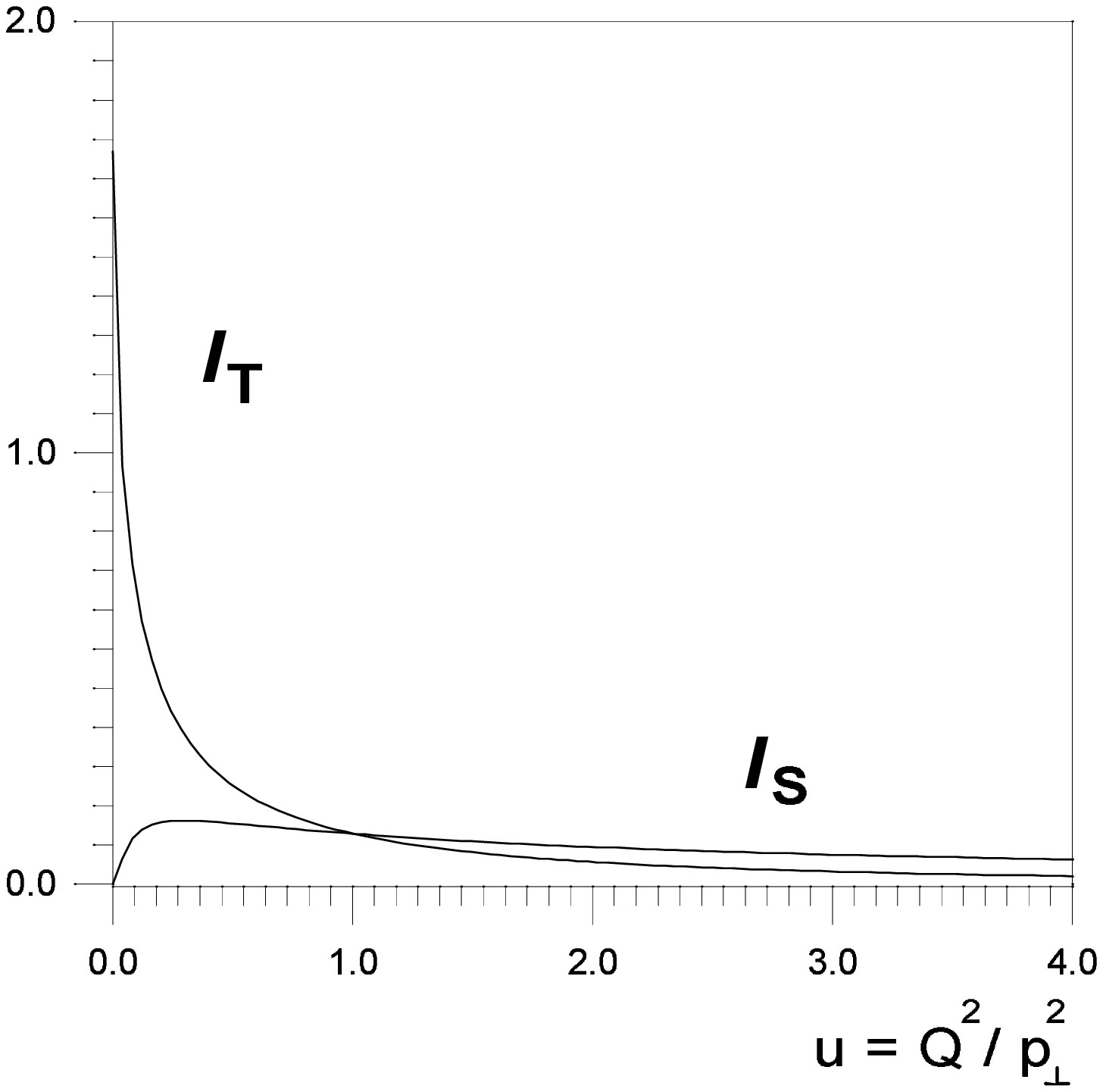}}
\vspace{-1.5cm}
\caption{Functions $I_T$ and $I_S$ for the process $\gamma ^*
q\to \rho^0 q$ or $\gamma ^* g\to \rho ^0 g$.}
\label{fig3}
\end{figure}

The dependence on photon virtuality is concentrated in factors
$I_{T},\; I_{S}$. The shapes of these functions depend weakly on
the form of wave function (value of quantity $b_V$). They are plotted
in Fig. \ref{fig3} for the $\rho ^0 $ meson production
($b_V=1.5$).

{\em If the virtuality of photon is less than \ptrs (or $u<1$)},
the amplitude for transverse photon dominates over longitudinal
one. Fig. \ref{fig3} shows very sharp peak in $I_T$ near $Q^2=
0$. It is due to items $\propto u\ln^2 u$ in eqs.  (\ref{19}),
(\ref{21}). The derivative of amplitude in $Q^2$ (in $u$)
diverges at $Q^2=0$, it is infrared unstable in contrast with
amplitude itself, which is infrared stable. The quantity $I_T$ is
reduced by half at $u\approx 0.1$. It means, that the scale of $Q^2$
dependence here is
\beq
\Lambda^2_{pert} \approx \ptrs/10.\label{lpert}
\eeq

The quantity $I_{S}(u)$ changes its sign at small enough $u=u_0$.
$u_0\approx 0.1$ for the asymptotical form of wave function
($b_V=0$); and $u_0\approx 0.02$ for the more wide wave function
with $b_V=1.5$. This behavior is similar to that for $J/\Psi$
production.

{\em If photon virtuality is large, $Q^2 > p_{\bot}^2$} (or $u>1$),
the amplitude with the scalar photon is dominant:
\beq
M_{\gamma ^*_S q\to V_L q} \propto {\ln{u} \over (Q^2)^{3/2}};
\;\;\;
M_{\gamma ^*_T q\to V_L q} \propto {p_{\bot} (\ln {u})^2 \over
(Q^2)^{2}}\quad \left(u={Q^2\over \ptrs} \right).\label{u>1}
\eeq

The pQCD cross sections for the light vector meson production on
gluon (the sum $d(\sigma _{\gamma ^* _T g\to V g} +d\sigma
_{\gamma ^* _S g\to V g})/ dp^2_{\bot}$ ) are presented in Fig.
\ref{fig3}. They are $s$-independent in the used first nontrivial pQCD
approximation.

\begin{figure}
\epsfxsize=15cm
\centerline{\epsffile{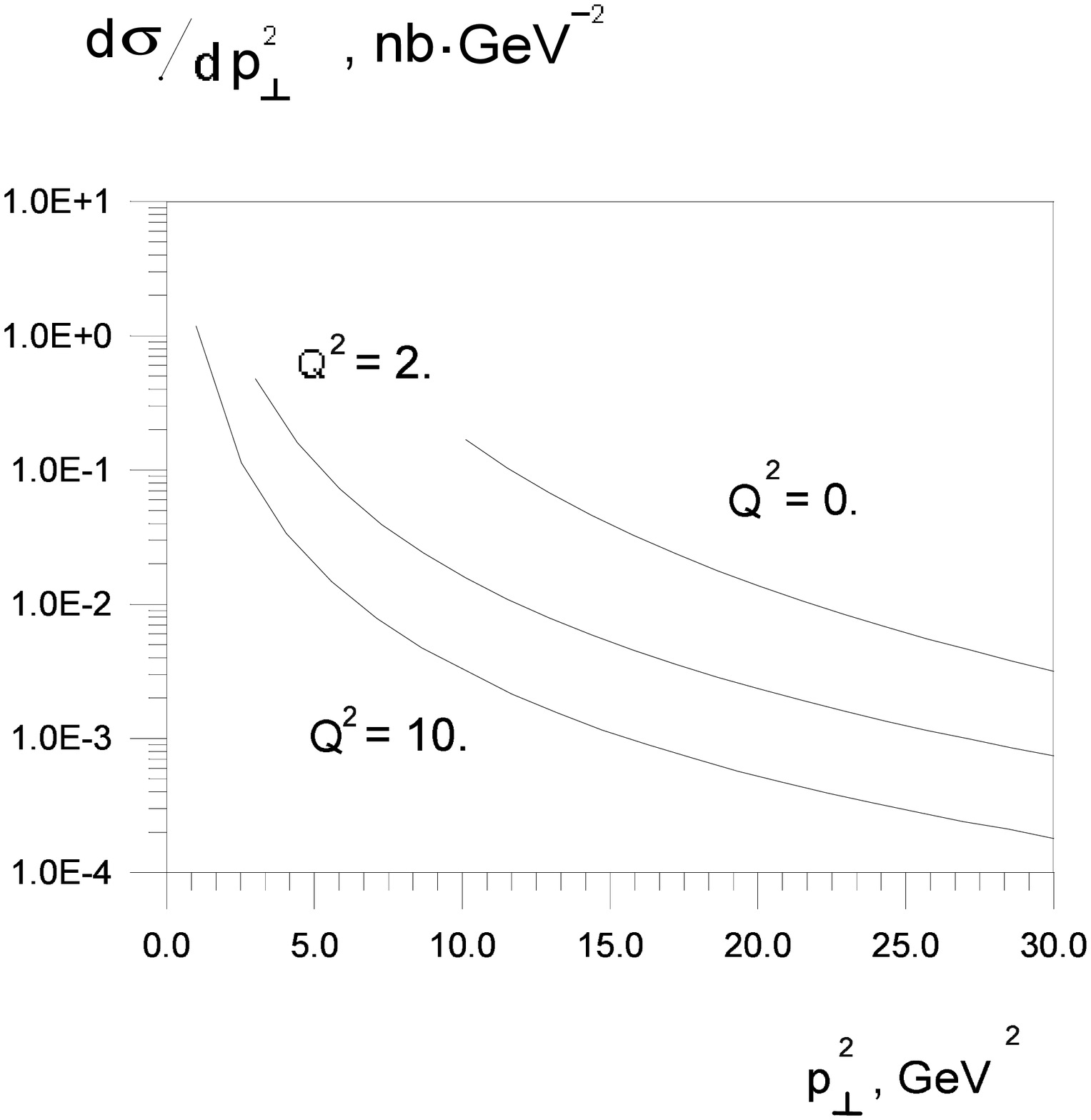}}
\vspace{-1.5cm}
\caption{Differential cross section of $\gamma^* g\to \rho^0 g$
process at $Q^2=0$, $Q^2=2\mbox{ GeV}^2$ and $Q^2= 10\mbox{
GeV}^2$.}
\label{fig4}
\end{figure}

{\large\em Some remarks related to $Q^2$ dependence.}

{\em (i)} The obtained scale of $Q^2$ dependence is substantially
lower than it was expected before calculations. Using of this
dependence will allow to improve the estimates for corresponding
production rate at $ep$ collisions.

{\em (ii)} Some features of $Q^2$ dependence for the high energy
asymptotic of LLA result were obtained in ref. \cite{Iv}. Here
the picture at small $Q^2$ (near mass shell) is similar to that
discussed above. Oppositely, far from mass shell (at $u\gg 1$)
the LLA amplitude contains an additional factor $\propto Q/\ptr$
in comparison with two--gluon approximation.

\section{Small \ptr limit for high $Q^2 \; (u<1)$ and coherence}

In the pQCD limit the amplitudes of photoproduction on quark or
gluon (\ref{PsiT},\ref{19},\ref{21}) diverge at $p_{\bot} \to 0$
(see \cite{GPS2} too). It means that "soft" nonperturbative
region $k_{\bot}\lessim\mu$ contributes substantially in this
range and pQCD calculations are infrared unstable. Therefore,
the details of \ptr dependence at $p_{\bot} \to 0$ are out of
range of the pQCD validity even at large $Q^2$ when we consider
production on color target.

The reason for this divergence is simple: in the above limit the
poles of both gluon propagators coincide, and we deal with the
integral
$$
\int J({\bf k}_{\bot}) d^2k_{\bot}/ k_{\bot}^4.
$$
In accordance with eq. (\ref{dipol}), $J \propto {\bf k}_{\bot}$
at ${\bf k}_{\bot} \to 0$ and $J \propto({\bf p-k})_{\bot}$
at $({\bf p-k})_{\bot}\to 0$. At ${\bf p}_{\bot}\to 0$ both
these zeroes coincide, and $J \propto k_{\bot}^2$. Then the
discussed divergence is logarithmic one only and the above
divergence is integrable, the total cross section is finite.

On the contrary, the $\gamma^*\gamma\to \Psi \Psi$ amplitude is
finite at $p_{\bot}\to 0$ due to additional factor $J\propto
k_{\bot}^2$ in the integrand. Therefore, soft part of integration
region gives a negligible contribution here.

The main difference between these amplitudes originates from the
fact that in the last case we deal with collision of two real
colorless objects; coherence between quarks results in the
additional suppression of soft nonperturbative contribution there
(\ref{dipol}).

The above comparison shows us that the coherence in both collided
particles should be taken into account to describe phenomena at
any $p_{\bot}$ within pQCD even in the region of large $Q^2$.

\section{The range of validity of pQCD results}

The above results show us that the using of pQCD for the
description of experimental data can be inaccurate in some
region of parameters. For example, the obtained scale of $Q^2$
dependence $\Lambda^2_{pert}\approx \ptrs/10$ (\ref{lpert}) is
smaller than the natural scale of this dependence near mass
shell $\Lambda^2_{soft} \approx m_{\rho}^2$ even at $\ptr =2.5$
GeV when our small parameter $\mu^2/\ptrs< 0.02$.

In this section we discuss the bounds of the pQCD validity
region in dependence of \ptr for {\em the photoproduction of
vector mesons consisting of light quarks}, like $\rho$ (provided
$s\gg\ptrs$). Below we use some single scale of QCD
nonperturbative effects (confinement, gluon correlations, etc.)
--- $\mu$. We will have in mind the value $\mu=0.2\div 0.3$ GeV,
which is close to the confinement scale, constituent quark mass,
etc.

In the discussion below we assume the impact representation to
be valid independent on validity of pQCD for description of
different factors in it. In particular, the proof of impact
representation in the lowest nontrivial approximation of pQCD is
valid even in the regions near the poles of quark propagators in
the impact factor, where its perturbative form (\ref{7},\ref{7a})
becomes incorrect.\\

{\large\em The model for amplitude near the bound of region of
pQCD validity.}

To study the bound of pQCD validity, we simulate nonperturbative
effects by adding of quantity $\mu^2$ (instead of $m^2$) in all
quark propagator denominators (assuming $\mu=200\div 300$
MeV)\footnote{ The nonperturbative effects in the vicinity of
poles of gluon propagators are suppressed due to property
(\ref{dipol}).}. Besides, we change the quantity $m$ from the
quark propagator nominator (in front of item $R$ in eq.
(\ref{5})) for some new quantity $A\sim\mu$:
\beq
{\bf P}(m)\to {\bf P}(\mu); \qquad mR(m)\to A\cdot R(\mu)\quad
(A\sim \mu).
\label{amu}
\eeq

The regions, where the amplitude is sensitive to value of $\mu$,
are beyond pQCD validity. {\bf\em We denote the bound ${\bf
p_{pert}}$ of pQCD validity region by relation}
\beq
d\sigma(\ptr\geq p_{pert}|\mu) > 0.5\cdot d\sigma(\ptr\geq
p_{pert}|\mu =0). \label{ppertdef}
\eeq

We begin with {\em the meson photoproduction by real (on shell)
photons}.

Contribution of item $R$ provides helicity conservation
(production of $V_T$) in the process. The item {\bf P} in the
impact factor gives longitudinal polarization of produced mesons
($V_L$). The contribution of this item decreases more slow with
\ptr due to extra power of momentum in nominator. Therefore,
this item gives amplitude at high enough \ptr.

Let us discuss the limit $\mu \ll\ptr$ in more detail. The
contribution of $R$ diverges in this limit due to integration
near the poles of quark propagators at $k_{\bot}=q_{i\bot}$, it
is $\sim \ln(\ptrs /\mu^2)$ (i.e. infrared unstable). It
dominates at not too large \ptr . Oppositely, the contribution
of ${\bf P}$ is finite in the discussed limit. It is infrared
stable, and it defines amplitude within the range of pQCD
validity.

Therefore, it is natural to assume that {\bf \bm the
contribution $P$ describes the point--like component of photon}
in the region where confinement effects are negligible. It
dominates at high values of \ptr and it provides production of
longitudinally polarized mesons. Similarly, {\bf \bm the
contribution $R$ for transverse photons describes the
hadron--like component of photon}. It dominates at not too high
values of \ptr and it provides helicity conservation here. In
addition to the boundary $p_{pert}$ (\ref{ppertdef}) we denote
boundary value ${\bf p_{hel}}$ by condition: At $\ptr> p_{hel}$
the mean helicity of produced V changes from transversal to
longitudinal one (i.e. hadron--like component $R$ becomes
relatively small).\\

{\large\bf\bm The bound of pQCD validity region, estimate of
$p_{pert}$}\\

We expect that $p_{hel}<p_{pert}$. Threfore, to find the bound
$p_{pert}$, one should consider point--like component of photon
(contribution {\bf P} in impact factor) only. We present two
estimate here.

{\em First estimate}. It is well known that the typical scale
of the $Q^2$ dependence for soft processes $\Lambda_{soft}^2
\approx m_{\rho}^2$ (here $m_{\rho}$ is the $\rho$ meson mass).
The known data shows us that this scale increases with $p_{\bot}$
growth.

The scale of $Q^2$ dependence obtained is $\Lambda^2_{pert}
\approx p_{\bot}^2/10$ for the $\rho$ photoproduction
(\ref{lpert}). The pQCD can be valid for description of the
discussed phenomena if only $\Lambda^2_{pert}>\Lambda^2_{soft}$,
i.e. at $\ptrs/ 10\; >\; m_{\rho}^2$ which leads to $\ptr
\gtrsim 3$ GeV. It does not contradict more refined estimate
below (\ref{ppert}).

\begin{figure}
\epsfxsize=15cm
\centerline{\epsffile{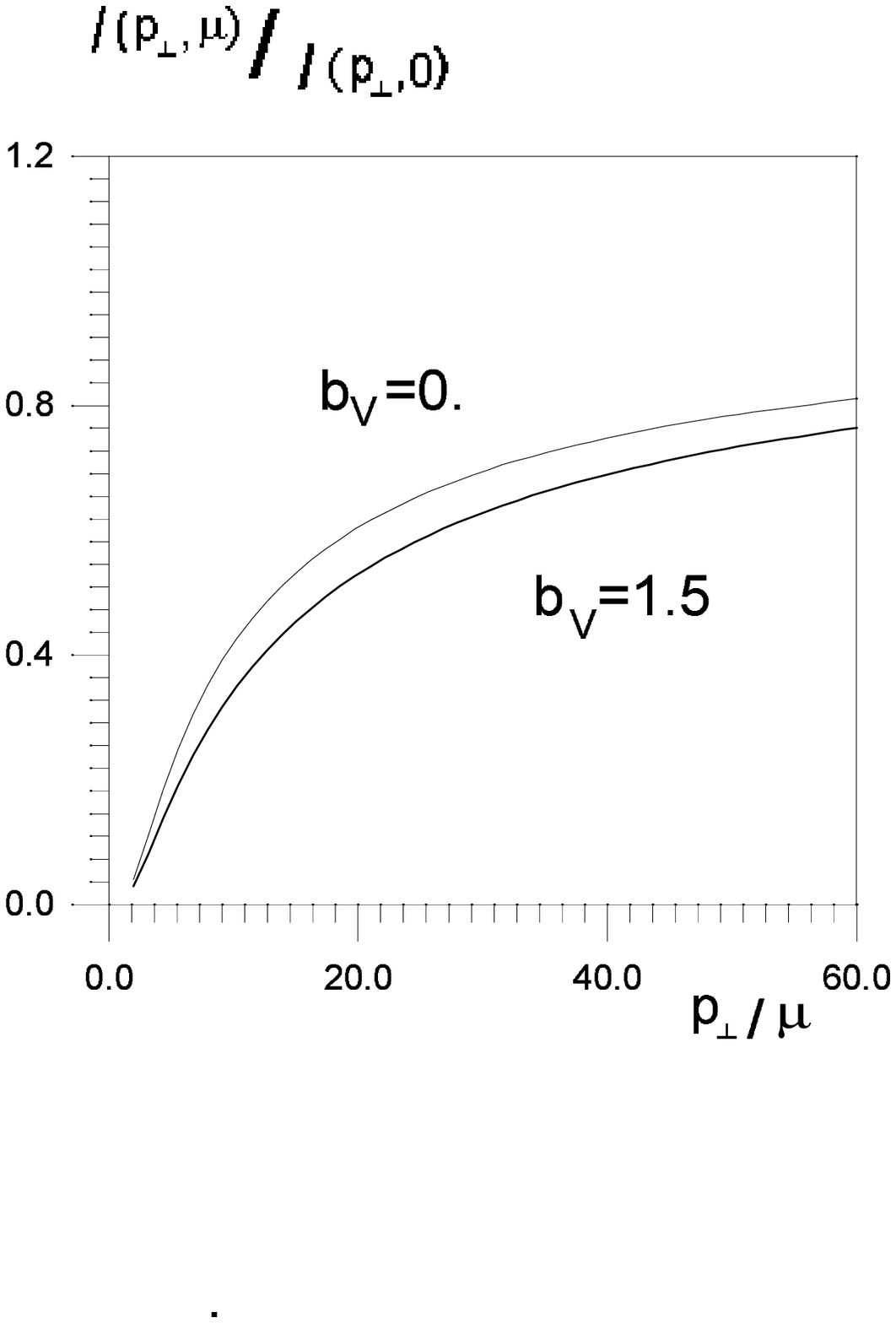}}
\vspace{-1.5cm}
\caption{The ratios $\Phi=M(\delta)/M(\delta =0)$ for mass shell
photons ($u=0$) in dependence on $\ptr/\mu=2\delta^{-1}$.}
\label{fig5}
\end{figure}

{\em Second estimate.} We calculated numerically the contribution
of item {\bf P} in impact factor (\ref{15})--(\ref{16}) with some
finite value of $\mu$ for different meson wave functions. Results
--- the ratios of $\Phi=M(\delta ,u)/ M(\delta =0, u)$), --- are
shown in Fig. \ref{fig5} for the important case of real photons ($u=0$).

Naturally, the ratio $\Phi\to 1$ at $(\ptr/\mu)\to \infty$.  That
is pQCD limit. We define value $p_{pert}$ by condition
$\Phi(p_{pert},\mu) =0.7$. At higher values of \ptr influence of
confinement effects for pQCD result in cross section is described
by factor $\Phi^2$ which is between 0.5 and 1.

It is seen that for mass shell photons $(p_{pert}/\mu)
\approx 30\div 40$. In this region the coefficient $b_V$ in the
$\rho$ meson wave function decreases up to $b_V\approx 0.7$.
Taking this fact into account, we have for $\mu =0.2$ GeV
\beq
p_{pert} \approx \left\{
\begin{array}{rlr}
7.5\mbox{ GeV}& \mbox{ for } b_V=0.7& (\rho - \mbox{ meson at }
\ptr\approx 7 \mbox { GeV}),\\
6.2\mbox{ GeV}& \mbox{ for } b_V=0 &(\phi - \mbox{ meson}).
\end{array}\right .\label{ppert}
\eeq
For $\mu=0.3$ GeV these quantities should be 1.5 times larger.

The obtained values of $p_{pert}$ (\ref{ppert}) for the mass shell
photons are very high. It is because the correction to the pQCD
result is governed by the parameter $\mu^2/\ptrs \ln^2(\ptrs
/\mu^2) $ but not the "natural" parameter $\mu^2/\ptrs$. The
effect of "$\mu$ corrections" in pQCD equations is enhanced near
the bounds of kinematical region, at $\xi\to \pm 1$. Therefore,
their influence is higher for the wave function, which is
"shifted" to these bounds (with $b_V>0$). <<It corresponds to the
table 1.>> In other words, the bounds for pQCD validity region
$p_{pert}$ are lower for the $\phi$ meson photoproduction
($b_V=0$) in comparison with that for $\rho$ photoproduction
($b_V=1.5$). The photon virtuality prevents quark propagators
from their poles while $\xi\neq\pm 1$. It is the reason why
$p_{pert}$ decreases fast with photon virtuality.
<<For example, >>
Our calculations show that
\beqn
p_{pert}^2& \approx& \left\{
\begin{array}{rr}
1.3 \mbox{ GeV}^2& \mbox { for } \rho \\
1 \mbox{ GeV}^2& \mbox { for } \phi \\
\end{array}\right\} \mbox { at } \mu =0.2\mbox{ GeV}, Q^2 =1
\mbox{ GeV}^2; \nonumber\\
p_{pert}^2& \approx& \left\{
\begin{array}{rr}
28 \mbox{ GeV}^2& \mbox { for } \rho \\
10 \mbox{ GeV}^2& \mbox { for } \phi \\
\end{array}\right\} \mbox { at } \mu =0.3\mbox{ GeV}, Q^2 =1
\mbox{ GeV}^2; \label{ppertq}\\
p_{pert}^2& \approx& \left\{
\begin{array}{rr}
3.3 \mbox{ GeV}^2& \mbox { for } \rho \\
2 \mbox{ GeV}^2& \mbox { for } \phi \\
\end{array}\right\} \mbox { at } \mu =0.3\mbox{ GeV}, Q^2 =2.25
\mbox{ GeV}^2; \nonumber\\
\eeqn

It is seen, that the the pure pQCD description with longitudinal
meson photoproduction become valid ealier in the $\phi$
photoproduction as compare with $\rho$ one.\\

{\large\bf \bm Signature of the pQCD validity for discussed
processes. Polarization of produced mesons. Estimate of
$p_{hel}$}\\

The above description shows that the signature of pQCD validity
is given by polarization of mesons consisting of light quarks and
the correct dependence on \ptr.

First, in the range of pQCD validity {\em these mesons should be
produced in the state with helicity 0 only}. This result takes
place for the production of both vector and tensor mesons
\cite{GPS1,GIv}. It is in strong contrast with the production in
"soft" region where helicity conservation takes place, and real
photons produce transversaly polarized mesons. (On the contrary,
at the production of mesons consisting of heavy quarks the
photon helicity transmits to meson in the main approximation.)

Second, the number of independent variables in the description of
cross section is reduced from three to two\footnote{ This
statement is valid for both discussed two--gluon approximation
and LLA. In the last case the dependence of the size of rapidity
gap $\eta$ corresponds to the perturbative Pomeron.}:
\beq
\ptr^6\fr{d\sigma}{d\ptrs} = F(\eta ,u);\quad \eta=\ln(s/\ptrs),\;
u=\fr{Q^2}{\ptrs}. \label{fin}
\eeq

Besides, the striking feature of results obtained is the very
{\em sharp dependence on photon virtuality near $Q^2=0$} (more
precise, on ratio $u$ for reactions (\ref{0}). The observation of
such a behavior will be a good additional test of pQCD.\\

Next point is to see for the crossover point, in which the
longitudinal polarization become dominant for the transversal
initial photon (the boundary $p_{hel}$). This boundary is below
boundary $p_{pert}$. Therefore, the calculations near this point
depend on detail of model more strong. To see qualitive features
of this crossover, the model (\ref{amu}) is used, in which
we fix coefficient $A=1$ GeV for definiteness.

\begin{figure}
\epsfxsize=15cm
\centerline{\epsffile{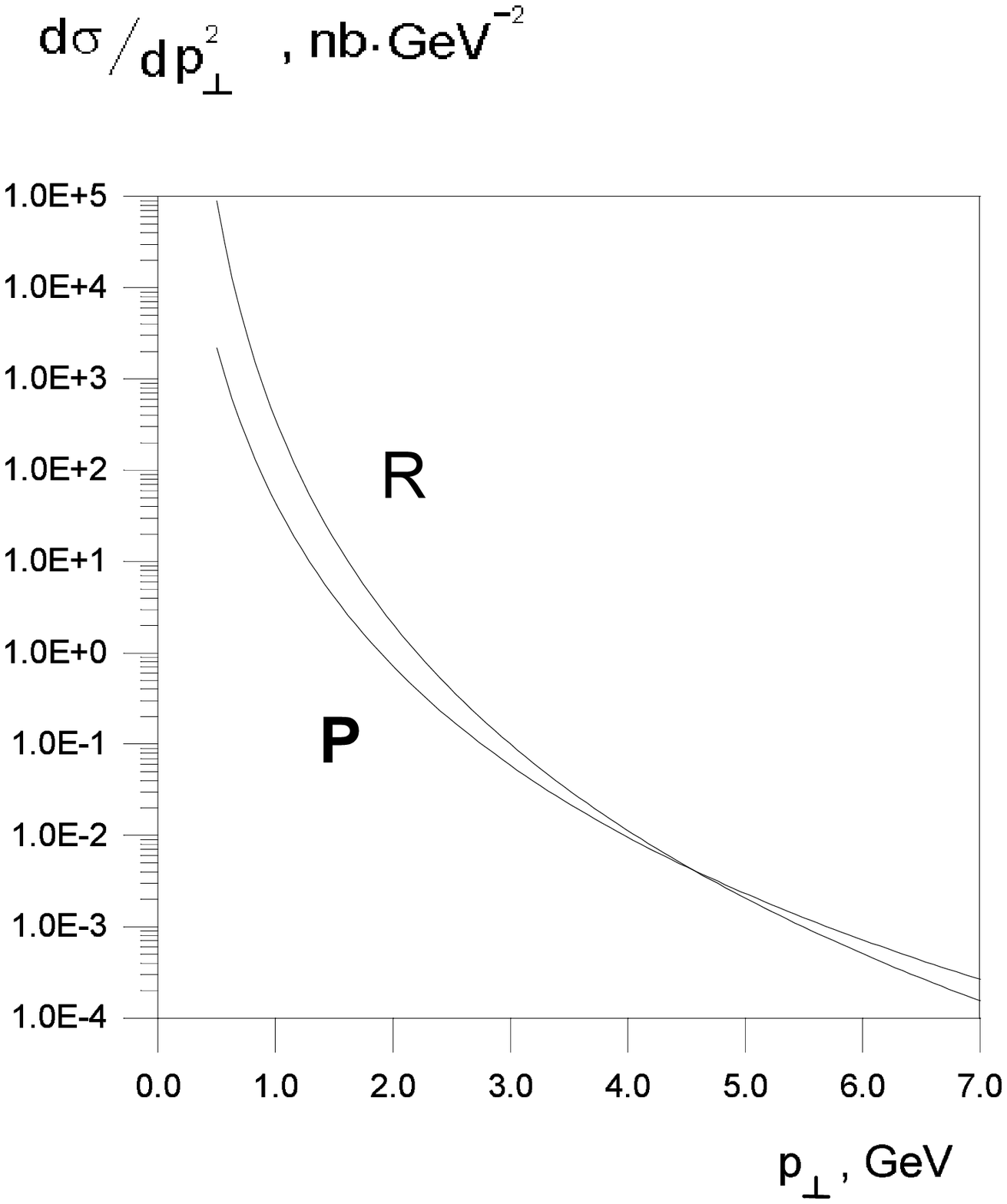}}
\vspace{-1.5cm}
\caption{The diffrerential cross sections of
$\rho^0$ meson  photoproduction ($b_V=1.5$) by
real photons for $\mu=200$ MeV. In these figures curves R
correspond to the production of transverse mesons (helicity
conserved contribution, item $R$, hadron--like
component of photon) and curves P correspond to the production
of longitudinal mesons (item ${\bf P}$, point--like
component of photon).}
\label{fig6a}
\end{figure}

\begin{figure}
\epsfxsize=15cm
\centerline{\epsffile{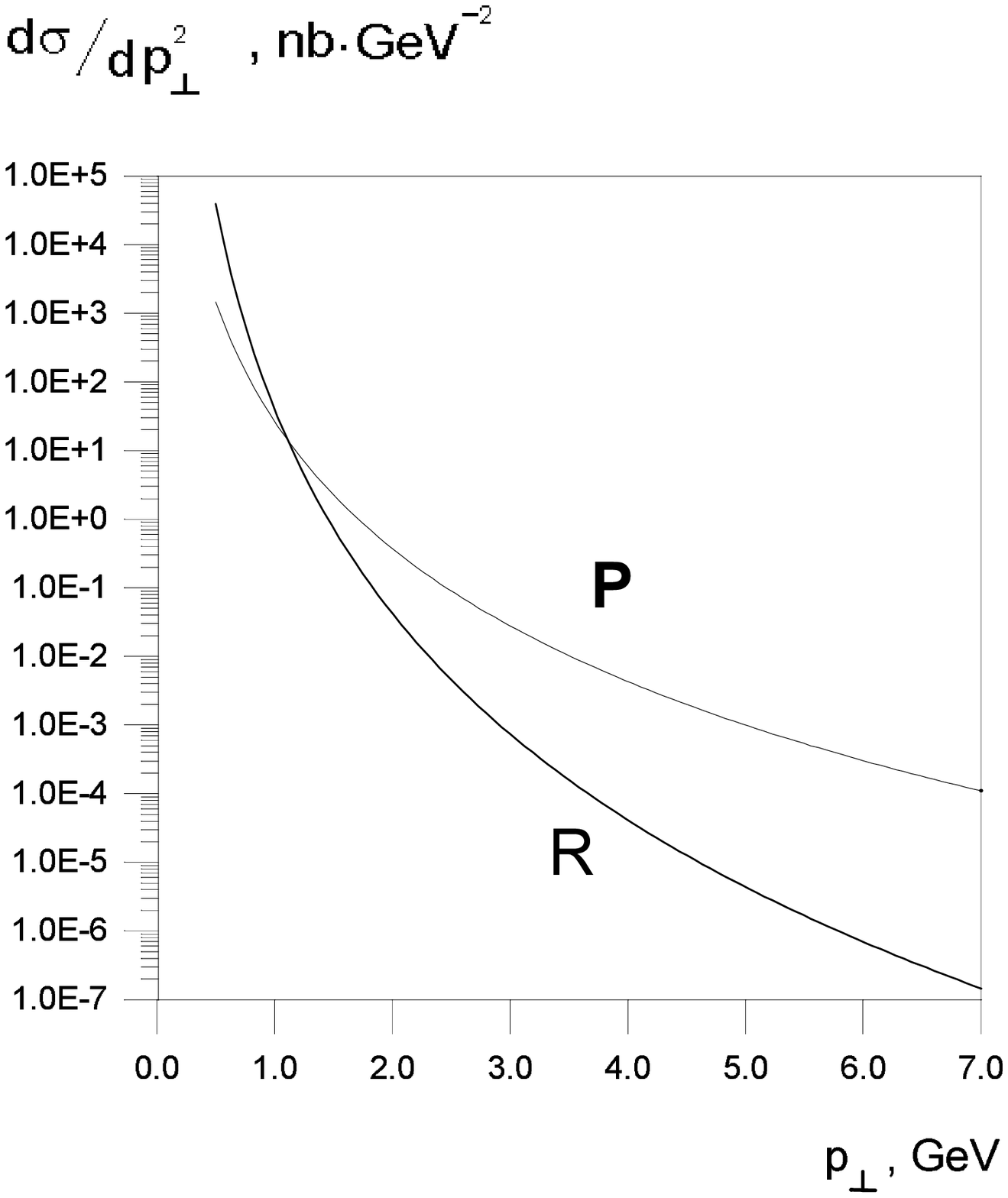}}
\vspace{-1.5cm}
\caption{The same figure as previous one, but for photoproduction
of $\phi$ meson ($b_V=0.0$) }
\label{fig6b}
\end{figure}

Figs. \ref{fig6a},\ref{fig6b}
shows the cross sections of photoproduction by
real photons for $\mu=200$ MeV. In these figures curves R
correspond to the production of transverse mesons (helicity
conserved contribution, item $R$ (\ref{7a}), hadron--like
component of photon) and curves P correspond to the production
of longitudinal mesons (item ${\bf P}$ (\ref{7}), point--like
component of photon). Fig. \ref{fig6a} shows curves for the
$\rho^0$ meson production ($b_V=1.5$). Fig. \ref{fig6b} shows
curves for the $\phi$ meson production (asymptotical wave
function, $b_V=0$).

In both cases the crossover point $p_{hel}$ is $1.5\div 5$ GeV. Next, the
admixture of transversaly polarized mesons at $\ptr>p_{hel}$ for
the $\rho$ photoproduction decriases with \ptr faster than that
for $\phi$ mesons. We expect, that this feature conserves for
virtual photons. It means, that in the data averaged over some
\ptr interval the fraction of longitudinal $\phi$'s is larger than
that for $\rho$'s. This conclusion is supported by data
\cite{exp2}.

\section{Brief discussion about some related papers}

All known for us papers, which treat the similar problems,
contain the essential phenomenological components (usually pQCD
inspired).

These models are based, in fact, on the impact representation
like (\ref{4}). The description begins with the diffractive
region (small \ptr). It is one of reasons why authors uses the
hadron--like component of photon (item $R$ but no point--like one
${\bf P}$) only with some parameter $\mu$ for detail description
of cross section. Therefore, these models predict the helicity
conservation in reaction $\gamma q\to\rho^0 q$. They don't
predict change of polarization of produced mesons at high
\ptr.

The quasi-elastic process $\gamma^*p\to\rho p$ (without proton's
dissociation) was studied in refs. \cite{Land,Cud}. In these
papers it was used the QCD inspired phenomenological model, which
can be represented as impact representation (\ref{4})
with the replacement of pQCD gluon propagators on the reggeized
ones. The $Q^2$ dependence for the forward scattering in this
model differs from that obtained in pQCD \cite{BrF}.

Recently the papers \cite{BrF,Rys,Nic} were published, where the
problems are studied that are close to those discussed above. In
these papers quasi-elastic photoproduction of vector mesons on
proton without proton's dissociation ($\gamma^* p\to Vp$) is
studied (with $V=\rho $ in \cite{BrF,Nic} and $V=J/\Psi$ in
\cite{Rys}).

The first stage in these papers corresponds to the simplest pQCD
diagram just as in our paper. The following stages are used some
features of processes at $p_{\bot}\approx 0$. To describe the
picture at large $p_{\bot}$ some phenomenological assumptions
were added in papers \cite{Rys,Nic}.

The $\rho$ meson photoproduction at $p_{\bot}\approx 0$ was
studied in ref. \cite{BrF}. The same very region for the
$J/\Psi$ photoproduction is the starting point for ref.
\cite{Rys}. Just in this case some features of LLA provides an
opportunity to use unitarity for construction of cross section
in terms of the LLA proton's gluon distribution\footnote{ This
basic construction is broken up at $\ptr\neq 0$. To describe the
$J/\Psi$ production in this region, it is used the additional
assumption in ref. \cite{Rys} that the object, which was the
proton's gluon distribution at $\ptr=0$, transforms to the
product of this distribution and some proton form factor
dependent on $p_{\bot}^2$ only.}. We consider quite other
kinematical region (\ref{range}).

The papers \cite{Nic} treat the process (\ref{0}). The crucial
point here is using of hadron--like component of photon (factor
$R$). It is the reason why these authors obtain transversal
polarization of $\rho$ meson for on shell photoproduction even
at large enough $p_{\bot}$.

{\large\em The photoproduction of pseudoscalar or tensor mesons
--- processes with three gluon exchange in the $t$ channel} (like
$\gamma^* q \to\pi^0 q$) relates to the Odderon problem. For the
mass shell photons these processes were considered in the similar
approach in ref. \cite{GIv}.

The impact representation like (\ref{4}) with three gluon
denominators and impact factors similar to those in \eq{14}
describe these processes \cite{GIv}).

These processes with virtual photon were also calculated. Here
the last integrations were numerical ones \cite{Iv2}. The
results obtained are similar to those for discussed case. The
new curve $I_T$ is similar to that in Fig. \ref{fig3}, but they are more
sharp at small $u$ as compare with the 2-gluon exchange (Fig. \ref{fig3}).

Note that the photoproduction of scalar (or tensor) meson on a
gluon is forbidden due to C parity conservation \cite{Gin}.
Therefore, the comparative study of the vector and scalar (or
tensor) meson production in $ep$ collision can give an
additional information about a gluon content of proton and
shadowing effects at small $x$.

\section{ Concluding remarks}

Let us summarize our predictions (mainly for HERA experiments)
related to the photoproduction of mesons consisting of light
quarks.

\begin{enumerate}
\item For the real photoproduction we expect change of mean
polarization of produced vector mesons at $\ptr\sim 1.5\div 5$
GeV.  Above this bound produced vector mesons should be mainly
longitudinally polarized. We expect that for the $\phi$
photoproduction this bound is lower than that for $\rho$, and
the fraction of transversal $\phi$ decreases with
\ptr more fast.

\item The pure pQCD regim is hardly observable for real photons
since the corresponding boundary values are very high, even for
$\mu =0.2$ GeV (\ref{ppert}). This regim can be seen better
in photoproduction by virtual photons (\ref{ppertq}). The
signatures for this regim are:
\begin{itemize}
\item Mesons are polarized longitudinally.
\item The number of independent variables is reduced up to
two in the description of quantity $\ptr^6(d\sigma/d\ptrs)$
(\ref{fin}).
\end{itemize}
\item One can consider the special region of large enough
\ptr (within the region of pQCD validity) and not too high
values of rapidity gap (say, $y<3$). In accordance with the
results of refs. \cite{Rys,Iv}, we expect that in this region
our two gluon approximation works good, i.e. the $y$--dependence
is weak and $u$--dependence is given by eqs. (\ref{19},\ref{21}).
\end{enumerate}

The photoproduction of jets in the "direct" configuration and
with the rapidity gap provides opportunity to see the same
mechanisms in processes with larger cross sections. First data
in this problem were reported recently \cite{exp3}. The results
of corresponding calculations are rather bulky and needs for
detail discussions. One can expect that the point $p_{pert}$
will be lower here than that for the vector meson production
(\ref{ppert}). \\

We are grateful to P.~Aurhence, A.C.~Bawa, W.~Buchmuller,
V.~Chernyak, R.~Cudell, A.~Efremov, A.~Grozin, A.~Kotikov,
L.~Lipatov, K.~Melnikov, M.~Ryskin and A.~Vainshtein for useful
discussions. This work is supported by grants ISF and Russian
Ministery of Science RPL300 and INTAS -- 93 -- 1180.


\begin{thebibliography}{99}

\bibitem{GIS2} I.F.~Ginzburg, D.Yu.~Ivanov, V.G.~Serbo. Preprint
TPI-Minn-94/14 --T// Theor. Phys. Inst. Univ. of Minnesota
(1994).

\bibitem{Land}
A.~Donnachie and P.~Landshoff, Phys. Lett. {\bf B185} (1987) 403;
Nucl. Phys. {\bf B311} (1989) 509

\bibitem{Cud}
J.R.~Cudell, Nucl. Phys. {\bf B336} (1990) 1

\bibitem{Nic}
B.Z.~Kopeliovich, J.~Nemchick, N.N~Nikolaev and B.G.~Zakharov,
Phys. Lett. {\bf B309} (1993) 179; {\bf 324} (1994) 469.


\bibitem{BrF}
S.J.~Brodsky, L.~Frankfurt, F.J.~Gunion, A.H.~Mueller and
M.~Strikman, Phys. Rev. D50 (1994) 3134.

\bibitem{Rys}
M.G.~Ryskin, Z. Phys. {\bf C57} (1993) 89; J.R.~Forshaw,
M.G.~Ryskin. DESY 94-162/ RAL-94-058(1994).

\bibitem{GPS1}
I.F.~Ginzburg, S.L.~Panfil and V.G.~Serbo, Nucl.Phys. {\bf B284}
(1987) 685

\bibitem{GPS2} I.F.~Ginzburg, S.L.~Panfil and V.G.~Serbo,
Nucl.Phys. {\bf B296} (1988) 569

\bibitem{GIS1} I.F.~Ginzburg, D.Yu.~Ivanov, V.G.~Serbo.
Sov.Yad.Fiz. {\bf 56} (1993) 45--56.

\bibitem{GISM}
M.N.~Dubinin, I.F.~Ginzburg, D.Yu.~Ivanov and V.G.~Serbo, in
preparation

\bibitem{Iv} D.Yu.~Ivanov, hep-ph/9508319, submitted to Phys. Rev. D.

\bibitem{theonew} A.~Donnachie and P.~Landshoff, report at
Photon'95 Workshop, Sheffield (1995); T.~Sjostrand and
G.~Schuler, report at Photon'95 Workshop, Sheffield (1995);
M.~Genovese, N.N~Nikolaev and B.G.~Zakharov, preprint CERN-TH
95/13 (1995).

\bibitem{exp} CHIO, W.D.Schambroom et al., Phys.Rev {\bf D26} (1982);
H1 Collab., S.Levonian, in Proc. XXVIII Recontre de Moriond, Les
Arcs, France, (1993) 529; ZEUS Collab., M.Costa, presented at the
DIS Workshop, Eilat, Izrael (1994); E665, C.Y.Fang,
FERMILAB--Conf 93/305 (1993); EMC, J.Ashman et al., Z. Phys. {\bf
C39} (1988) 169.

\bibitem{exp2} M.Arneodo et. al. (NMC Collab), Nucl. Phys.
{\bf B429} (1994) 503.

 \bibitem{exp3} ZEUS collab., preprints DESY 94-198; 94-210;
95-093; reports of E.~Barberies, G.~Glasman, M.~Arneodo,
S.~Kartik, L.~Sinclair at Photon'95 Workshop, Sheffield (1995);
H1 collab., report of A.~Rostovtsev at Photon'95 Workshop,
Sheffield (1995).

\bibitem{GKST} I.F.~Ginzburg,  G.L.~Kotkin, V.G.~Serbo,
V.I.~Telnov, Sov. Pis`ma ZhETF {\bf 34} (1981) 514; Nucl. Instr.
Methods 205 (1983) 47.

\bibitem{I} N.~Isgur and C.H.~Llewellyn Smith, Phys. Rev. Lett.
{\bf 52} (1984) 1080; Phys. Lett. {\bf B217} (1989) 535

\bibitem{Br} S.J.~Brodsky in Proc. 9 Intern. Workshop on
Photon--Photon Collisions San Diego, USA, World Sc.(1992) p.209.

\bibitem{GIv} I.F.~Ginzburg and D.Yu.~Ivanov, Nucl. Phys. B
(Proc.  Suppl.) {\bf 25B} (1992) 224; Nucl.Phys. {\bf B388}
(1992) 376

\bibitem{LFr}
L.N.~Lipatov and G.V.~Frolov, Sov. Yad. Fiz. {\bf 13} (1971) 588

\bibitem{ChWu}
H.~Cheng and T.T.~Wu, Phys. Rev. {\bf D1} (1970) 3414


\bibitem{Lip} V.S.~Fadin, E.A.~Kuraev, L.N.~Lipatov. Sov. Phys.
JETP {\bf 45} (1977) 199;\ Ya.Ya.~Balitski, L.N.~Lipatov. Sov. J.
Nucl.Phys. {\bf 28} (1978) 822;\ L.N.Lipatov. Sov. Phys. JETP
{\bf 63} (1986) 904;\ in "Perturbative QCD". ed. A.H.Mueller,
World Sc., Singapore, (1989); M.F.~McDermott, J.R.~Forshow and
G.G.~Ross, preprint CERN-th/95-4 (1995).

\bibitem{MT} A.H.~Mueller, W.K.Tang. Phys. Lett. {\bf B284} (1993) 123.


\bibitem{LeBr}
G.P.~Lepage, S.J.~Brodsky, Phys. Rev. {\bf D22} (1980) 2157

\bibitem{ChAZh}
V.L.~Chernyak and A.R.~Zhitnitsky, Phys. Rep. {\bf 112} (1984)
173

\bibitem{BaGr}
V.N.~Baier and A.G.~Grozin, Fiz. Elem. Chast. i Yadra (Sov.
Journ. of Particles and Nuclei) {\bf 16} (1985) 5


\bibitem{Gin}
I.F.~Ginzburg, JETP Lett. {\bf 59} (1994) 579.

\bibitem{Iv2} D.Yu.~Ivanov. In preparation.

\end{thebibliography}
\end{document}